\documentclass[conference]{IEEEtran}
\IEEEoverridecommandlockouts
\usepackage[utf8]{inputenc}
\usepackage{amssymb}
\usepackage{amsmath}
\usepackage{graphicx}
\usepackage{nomencl}
\usepackage{etoolbox}
\usepackage{algorithm} 
\usepackage{algpseudocode} 
\usepackage{nomencl}
\usepackage[T1]{fontenc}
\usepackage{cite}
\usepackage{array,multirow,graphicx}
\usepackage{multirow}
\usepackage{mathtools}
\usepackage{colortbl}

\def\BibTeX{{\rm B\kern-.05em{\sc i\kern-.025em b}\kern-.08em
    T\kern-.1667em\lower.7ex\hbox{E}\kern-.125emX}}
\begin{document}
\title{Hybrid Quantum-Classical Unit Commitment\\
\thanks{ This work was supported by National Science Foundation under Grant ECCS-1944752.

%R. Mahroo and A. Kargarian are with the Department of Electrical and Computer Engineering, Louisiana State University, Baton Rouge, LA 70803 USA (e-mail: rmahro1@lsu.edu, kargarian@lsu.edu).}
}}
\author{\IEEEauthorblockN{Reza Mahroo}
\IEEEauthorblockA{\textit{Electrical and Computer Engineering Department} \\
\textit{Louisiana State University}\\
Baton Rouge, US \\
rmahro1@lsu.edu}
\and
\IEEEauthorblockN{ Amin Kargarian}
\IEEEauthorblockA{\textit{Electrical and Computer Engineering Department} \\
\textit{Louisiana State University}\\
Baton Rouge, US \\
kargarian@lsu.edu}

}
%\title{Hybrid Quantum-Classical Unit Commitment}
%\author{Reza Mahroo, \textit{Student Member, IEEE}, Amin Kargarian, \textit{Senior Member, IEEE}
%\date{December 2021}
%\thanks{ This work was supported by National Science Foundation under Grant ECCS-1944752. 

%R. Mahroo and A. Kargarian are with the Department of Electrical and Computer Engineering, Louisiana State University, Baton Rouge, LA 70803 USA (e-mail: rmahro1@lsu.edu, kargarian@lsu.edu).}}
\maketitle
\begin{abstract}
    This paper proposes a hybrid quantum-classical algorithm to solve a fundamental power system problem called unit commitment (UC). The UC problem is decomposed into a quadratic subproblem, a quadratic unconstrained binary optimization (QUBO) subproblem, and an unconstrained quadratic subproblem. A classical optimization solver solves the first and third subproblems, while the QUBO subproblem is solved by a quantum algorithm called quantum approximate optimization algorithm (QAOA). The three subproblems are then coordinated iteratively using a three-block alternating direction method of multipliers algorithm. Using Qiskit on the IBM Q system as the simulation environment, simulation results demonstrate the validity of the proposed algorithm to solve the UC problem.
\end{abstract}
\vspace{12pt}
%%%%%%%%%%%%%%%%%%%%%%%%%%%
%%%%%%%%%%%%%%%%%%%%%%%%%%%%%%%%%%%%%%%%%%%%%%%%%%%%%%%%%%%%%
\section{Introduction}
Computing plays a pivotal role in power system modeling and analysis. As the mathematical challenges of real-world problems increase, progress in advanced computing technologies becomes more crucial \cite{eskandarpour2020quantum}. Though still in the early stages, algorithms performed on quantum computers promise to complete important computing tasks faster than they could ever be done on traditional computers \cite{chen2019hybrid}. Despite substantial studies on quantum computing applications in a wide range of areas, its application to power systems has largely remained intact. This paper aims to perform a hybrid quantum-classical algorithm to solve the unit commitment (UC) problem, a computationally expensive problem for conventional computers.
%The proposed algorithm relies on decomposing the UC problem, a mixed-integer nonlinear problem (MINLP)

UC is a fundamental optimization problem in power systems operation. It aims to determine the commitment of generating units to supply the demand and meet technical constraints in a cost-effective manner. The UC problem is generally cast as a mixed-integer nonlinear programming (MINLP), which is expensive to solve using classical solvers. Solving advanced UC problems efficiently in a reasonable amount of time is of paramount importance \cite{abujarad2017recent}. The computational burden exponentially increases as the number of generating units increases. Thus, developing effective approaches to handle the UC problem becomes crucial.

There are a variety of solvers to solve or approximate mixed-binary optimization problems with classical or heuristic approaches. DIscrete and Continuous OPTimizer (DICOPT) is a program that solves MINLP problems by solving a series of nonlinear programming and mixed-integer programming (MIP) problems \cite{grossmann2002gams}. BARON is a popular option for solving MINLP problems \cite{sahinidis1996baron}. This solver relies on solving a series of convex underestimating subproblems arising from the evolutionary subsection of the search area. IBM’s CPLEX is a famous mathematical solver that works based on branch-and-cut to find exact or approximate MIP solutions \cite{KargarianWeb}. 

Solving optimization problems using quantum computers is mainly restricted to quadratic unconstrained binary optimization (QUBO) problems \cite{zahedinejad2017combinatorial}. A few algorithms, such as quantum approximate optimization algorithm (QAOA) \cite{farhi2014quantum} and variational quantum eigensolver (VQE) \cite{peruzzo2014variational}, are available to solve QUBOs. Given the mixed-integer nature of many practical optimization problems, we need to extend the quantum optimization techniques to cope with MINLP problems on current quantum devices. In this regard, to solve the UC problem, the authors in \cite{ajagekar2019quantum} have proposed two modifications to enable using quantum algorithms:
\begin{itemize}
    \item Adding slack variables to convert inequality constraints into soft equality constraints
    \item Discretizing continuous variables to h partitions to have pure binary variables
\end{itemize}

The main drawback is that for solving a UC problem with N units, N(h+1) qubits are required, which is not practical. 

In this paper, a hybrid quantum-classical algorithm is proposed to solve the UC problems. We first decompose UC into three subproblems, namely a quadratic subproblem, a QUBO, and a quadratic unconstrained subproblem. The QUBO subproblem is solved using a quantum QAOA algorithm, and the other two subproblems are handled by a classical solver. We then use a variant of alternating direction method of multiplier (ADMM) \cite{boyd2011distributed} to coordinate the QUBO and non-QUBOs (referring to quadratic and quadratic unconstrained subproblems) iteratively. Although the standard ADMM was originally developed for solving convex problems, recent studies  \cite{sun2019two, jiang2019structured, gambella2020multiblock} have developed heuristic ADMMs that can be applied to a variety of nonconvex problems. We have used one of these variants, namely a three-block ADMM. Simulations are carried out using Qiskit on the IBM Q system, and the validity of the proposed algorithm is studied.
%%%%%%%%%%%%%%%%%%%%%%%%%%%%%%%%%%%%%%%%%%%%%%%%%%%%%%%%%%
\section{Unit Commitment}
The UC formulation is a large-scale mixed-integer problem. Solving UC is computationally expensive because of generating unit nonlinear cost functions and the combinatorial nature of its set of feasible solutions. The UC solution determines the combination of units’ on/off status to meet the load at a time. UC a single time period is formulated as:
\begin{equation} \tag{1a}
\label{1a}
\min_{y_i,p_i} \sum_{i=1}^N f_i = \sum_{i=1}^N \left(A_{i} y_i  + B_i p_i + C_i p_i^2  \right)
\end{equation}
s.t. 
\begin{equation} \tag{1b}
\label{1b}
\sum_{i=1}^N p_i = L ,
\end{equation}
\begin{equation} \tag{1c}
\label{1c}
P_i^{min} y_i \leq p_i \leq P_i^{max} y_i ; \mbox{  } \forall i,
\end{equation}
\begin{equation} \tag{1d}
\label{1d}
y_i \in \{0,1\};  \mbox{  } \forall i.
\end{equation}
where $A_i$, $B_i$, and $C_i$ are the fixed cost coefficients of unit $i$. $y_i$ is a binary variable which is 1 if unit $i$ is on and 0 otherwise. $p_i$ is the power generated by unit $i$. Constraint (1b) preserves the system power balance. $L$ is the total load. Constraint (1c) limits the generated power of unit $i$ to $P_i^{min}$ and $P_i^{max}$. Note that the UC problem is further subject to more operational constraints like spinning reserve, ramping up/down limits, minimum on and off time constraints, and network constraints. For simplicity, these constraints are not considered in this paper.

\section{Three-Block Decomposition}
We decompose UC problem (1) into three subproblems, including a QUBO and two non-QUBOs. In the first step of the UC reformulation, we relax binary variable $y_i$ to vary continuously as $0 \leq y_i \leq 1$. We then introduce a new auxiliary binary variable $z_i$, a new auxiliary continuous variable $r_i$, and a new constraint (2c) to preserve the characteristics of problem (1). 
 \begin{equation} \tag{2a}
\label{2a}
\min_{y_i,p_i,z_i,r_i} \sum_{i=1}^N f_i 
\end{equation}
\centerline{s.t. (\ref{1b})-(\ref{1c})}
\begin{equation} \tag{2b}
\label{2b}
0 \leq y_i \leq 1;  \mbox{  } \forall i
\end{equation}
\begin{equation} \tag{2c}
\label{2c}
y_i - z_i + r_i = 0 : \lambda_i ;  \mbox{  } \forall i
\end{equation}
\begin{equation} \tag{2d}
\label{2d}
r_i = 0;  \mbox{  } \forall i
\end{equation}
\begin{equation} \tag{2e}
\label{2e}
z_i \in \{0,1\};  \mbox{  } \forall i,
\end{equation}
where $\lambda_i$ is the dual variable pertains to constraint (2c). Now, if we divide the variables into three sets \( \Delta = \{y_i,p_i\}\), $z_i$, and $r_i$, then relax constraints (\ref{2c}), the objective function and remaining constraints have a separable structure with respect to each variable set. The augmented Lagrangian of problem (2) is defined as:
\begin{equation} \tag{3a}
\label{3a}
\begin{aligned}
   {} & \mathcal{L}_\rho(p_i,y_i,z_i,r_i,\lambda_i)=\sum_{i=1}^N f_i + \frac{\beta}{2}||r_i||^2_2   \\
   & +\sum_{i=1}^N \lambda_i(y_i-z_i+r_i) +\frac{\rho}{2}\sum_{i=1}^N||y_i-z_i+r_i||^2_2,\\
\end{aligned}
\end{equation}
\centerline{s.t. (\ref{1b})-(\ref{1c}), (\ref{2b}), and (\ref{2e}).} \\
To solve (3) in three separate blocks, we now implement a 3B-ADMM shown in Algorithm 1. The first block is a quadratic optimization with $\{y_i,p_i\}$ as variables and given $z_i$ and $r_i$ as constants received from the other two subproblems. If this relaxed UC problem is infeasible, then so is the UC problem (1), and Algorithm 1 can be terminated. In the second block, we have a QUBO problem with respect to auxiliary binary variables $z_i$ and given $\{y_i,p_i\}$ and $r_i$. Third block is a unconstrained quadratic problem over auxiliary variables $r_i$ and given $\{y_i,p_i\}$ and $z_i$. 3B-ADMM is a heuristic approach due to the nonconvexity of the second block QUBO problem. However, this algorithm is guaranteed to converge under some restricting condition for large enough $\rho>\beta$.

The assumptions under which Algorithm 1 converges to a stationary point of the augmented Lagrangian $\mathcal{L}_\rho$ \cite{gambella2020multiblock, wang2019global} are:
\begin{enumerate}
  \item (Feasibility). The original problem is feasible. This condition holds since for any fixed $y_i$ and $z_i$ there is a $r_i$ that satisfies this constraint.
  \item (Coercivity). The objective function is coercive over constraint (2c). $\frac{\beta}{2}||r_i||^2_2$ is a quadratic term, so it is coercive. Since other variables are bounded, coercivity holds for them.
  \item (Lipschitz subminimization paths). For two consecutive iterations and a constant $M$, it is possible to have the condition $||y_i^{(\nu-1)}-y_i^{(\nu)}|| \leq M||y_i^{(\nu-1)}-y_i^{(\nu)}||$ for variable $y_i$. This condition holds with a constant $M=1$. Note that this condition is true for variables $z_i$ and $r_i$ as well.
  \item (Objective regularity). Objective (2a) is a lower semi-continuous over constraints (1b)-(1c) and (2b). This condition holds since objective (2a) is a convex function, and the set of constraints (1b)-(1c) and (2b) is convex \cite{gambella2020multiblock}.
\end{enumerate}
Moreover, if $\mathcal{L}_\rho$ is a Kurdyka–Łojasiewicz (KŁ) function, \cite{bolte2007lojasiewicz, attouch2013convergence} it would converge globally. Since objective function (3a) is semi-algebraic, it is a KŁ function. Therefore, Algorithm 1 converges to a stationary point of $\mathcal{L}_\rho$ function with starting from any point and any large enough $\rho>\beta$. 
\begin{algorithm}
	\caption{3B-ADMM algorithm} 
	\begin{algorithmic}[1]
	\State Initialize: $\nu=1$, $\lambda_i^{(0)}$, $\rho>\beta>0$, $z_i^{(0)}, r_i^{(0)}, \epsilon >0$.
		\For {$\nu=1,2,\ldots ,$}
		   \State First block update:\\ $\Delta^{(\nu)} = \underset{p_i,y_i}{\operatorname{argmin}}\mathcal{L}_\rho(\Delta,z_i^{(\nu-1)},r_i^{(\nu-1)},\lambda_i^{(\nu-1)})$ 
		   \State s.t (1b)-(1c) and (2b).
		   \State Second block update:\\ $z_i^{(\nu)} = \underset{z_i \in \{0,1\}}{\operatorname{argmin}}\mathcal{L}_\rho(\Delta^\nu,z_i,r_i^{(\nu-1)},\lambda_i^{(\nu-1)})$.
		   \State Third block update:\\ $r_i^{(\nu)} = \underset{r_i}{\operatorname{argmin}}\mathcal{L}_\rho(\Delta^\nu,z_i^\nu,r_i,\lambda_i^{(\nu-1)})$.
		   \State Dual variable update:\\ $\lambda_i^{(\nu)} = \frac{\rho}{2} (y_i^{(\nu)}-z_i^{(\nu)}+r_i^{(\nu)})+\lambda_i^{(\nu-1)}$.
		   \If {$||y_i^{(\nu)}-z_i^{(\nu)}+r_i^{(\nu)}|| \leq \epsilon$}
		        {Stop}. \Else
		        \State $\nu \leftarrow \nu+1$.
		   \EndIf
		\EndFor
		\State Return $(y_i,p_i,z_i,r_i)$.
	\end{algorithmic} 
\end{algorithm}

By fixing the variables $r_i=0$, we will have a two-block ADMM and skip the third block update. However, having auxiliary variables $r_i$ has two main advantages. First, it guarantees the feasibility of the problem by ensuring that for all fixed variables $y_i$ and $z_i$, an $r_i$ would exist such that constraint (2c) would hold. Second, constraint (2d) can be handled independently of (2c) and add the convex term $\frac{\beta}{2}||r_i||^2_2$ to the objective function.
%Due to the penalty term in the augmented Lagrangian, the multiplier updating in step 11 of Algorithm 1 proceeded smoothly \cite{bertsekas2014constrained}.
%%%%%%%%%%%%%%%%%%%%%%%%%%%%%%%%%%%%%%%%%%%%%%%%%%%%%%%%%%%%%%%%%%%%%%%%%%%%
\section{Quantum Approximation Optimization Algorithm}
While the two non-QUBO subproblems can be solved using a classical computing solver, the QUBO is solved using QAOA, which is a quantum computing-based algorithm. QAOA intends to approximate QUBO problems, which is a kind of combinatorial optimization problem. Combinatorial optimization is the process of searching for minima (or maxima) of an objective function whose domain is a discrete but large configuration space. Here, we want to minimize QUBO function $C(z)$, where $z$ denotes the set of variables $\{z_1,z_2,...,z_n\}$.  $z_i$ values can be $+1$ or $-1$. Therefore, to meet the standard form of QAOA, we have to convert our binary variables $z_i \in \{0,1\}$ as follows to take $\{+1,-1\}$:

\begin{equation} \tag{4}
\label{4}
z_i \rightarrow 2z_i -1;  \mbox{  } \forall i.
\end{equation}

QAOA acts on $n$ quantum-bits (qubit), where each qubit represents the state of one of the binary variables. Initially, this algorithm begins with all of the qubits initialized at state $|0\rangle$. The next step is to put all qubits into an equal superposition by applying $H^{\otimes n}$, the Hadamard operator on each qubit. In the following, we aim to alter the amplitudes so that those with small $C(z)$ coefficients will grow and those with large $C(z)$ coefficients will decrease. Thus, there will be a greater chance of finding a bitstring with a small value of $C(z)$ when we measure the qubits. To this end, first, we apply the unitary operator, so-called cost Hamiltonian, $U(\gamma,C)=e^{i\pi \gamma C(Z)/2}$. $\gamma$ is a variational parameter that we adjust its value to achieve the best possible results. $C(Z)$ is a diagonal matrix in the computational basis that its elements correspond to all the possible values for the second block objective function. $C(Z)$ consists of  the Pauli-$Z$ operator on each qubit based on the argument $z$.

So far, the result of this algorithm is a summation of all possible bitstrings, with complex coefficients that depend on $C(z)$. Also, the probability of all bitstring is equal at this stage. Then we apply the second unitary operator, so-called mixing Hamiltonian, $U(\beta,B)=e^{i\pi \beta B/2}$, where $\beta$ is a variational parameter, $B=\sum_{i=1}^N X_i$, and $X$ is Pauli-$X$ operator. This operator rotates each qubit $\beta$ degree around the X-axis on the Bloch sphere. Unlike the previous operator, this is not a diagonal operator on the computational basis, and the probability of final states will not be the same for all bitstrings. This algorithm consists of repeating the first two steps $P$ times, where $P$ is the depth of the circuit regardless of how many qubits there are. The state of the qubits after all operators are applied to the initial bitstring is:
\[|\gamma,\beta \rangle = U(\beta_P,B)U(\gamma_P,C)...U(\beta_1,B)U(\gamma_1,C) |s^n\rangle.\]

To find the best variational parameters $\gamma$ and $\beta$, after preparing the state $|\gamma,\beta \rangle$ we use a classical technique to minimize the expectation value $F(\gamma,\beta)=\langle \gamma,\beta | C(Z) |\gamma,\beta \rangle$. 

\begin{figure}
\centering
    \includegraphics[width=.48\textwidth]{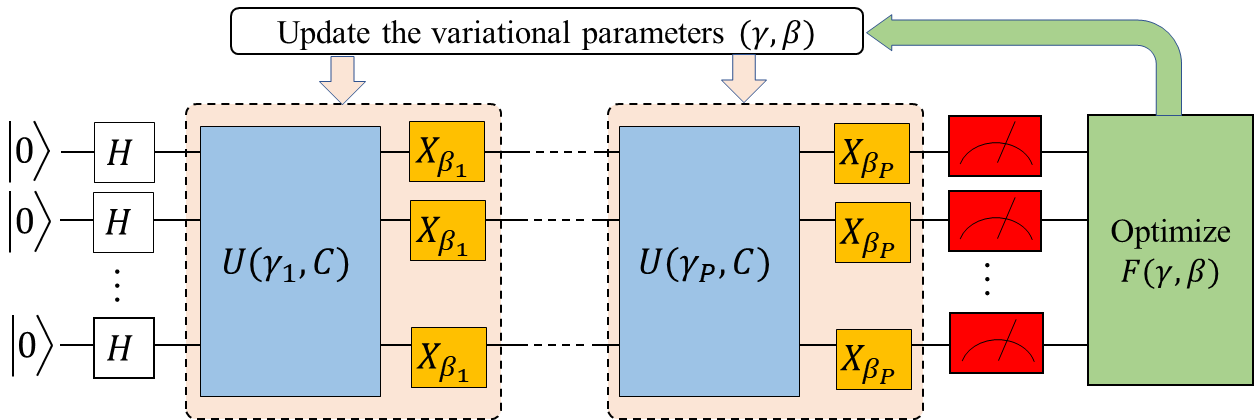}
\caption{QAOA circuit diagram.}
\label{fig1}
\end{figure}

Figure \ref{fig1} illustrates the prototype of a QAOA circuit. In a nutshell, we follow the steps below to implement QAOA:
\begin{enumerate}
    \item Prepare an initial state and apply Hadamard operator
    \item Define the cost Hamiltonian based on the QUBO objective function
    \item Define the mixing Hamiltonian.
    \item Construct the quantum circuits $U(\gamma,C)$ and $U(\beta,B)$, then repeat it $P$ times
    \item Calculate the optimal value of the variational parameters using a classical solver
    \item Measure the final state that reveals approximate solutions to the optimization problem
\end{enumerate}

%%%%%%%%%%%%%%%%%%%%%%%%%%%%%%%%%%%%%%%%%%%%%%%%%%%%%%%%%%%%%%%%%%%%%%%%%%%%
%%%%%%%%%%%%%%%%%%%%%%%%%%%%%%%%%%%%%%%%%%%%%%%%%%%%%%%%%%%%%%%%%%%%%%%%%%%%
\section{Case Studies}
As the base case, the centralized UC problem (1) is implemented in Python 3.9 using the Pyomo package \cite{bynum2021pyomo}, and solved using IPOPT solver \cite{wachter2006implementation}. The reformulated UC problem (3) is then solved using Algorithm 1 under two strategies:

$S1)$ solving all three blocks using the classical solver.

$S2)$ solving first and third blocks using classical solver, and the second block using QAOA algorithm. 

The quantum circuit for solving the QUBO problem is implemented in Python using the IBM Qiskit package \cite{Qiskit}. In the QAOA algorithm, the depth of the system $P$ is set to 2, and the maximum iteration is 100. A ten-generating unit power system is used. Table \ref{table:1} represents the system parameters, and Table \ref{table:2} depicts the base case results for a variety of load levels.

%%%%%%%%%%%%%%%%%%%%%%%%%%%%%%%%%%%%%%%%%%%%%%%%%%%%%%%%%%%
\begin{table*}[ht]
\caption{Generating Unit Parameters}
\label{table:1}
\centering
\begin{tabular}{p{0.06\linewidth}p{0.06\linewidth}p{0.06\linewidth}p{0.06\linewidth}p{0.06\linewidth}p{0.06\linewidth}p{0.06\linewidth}p{0.06\linewidth}p{0.06\linewidth}p{0.06\linewidth}p{0.06\linewidth}}
\hline
Unit i & 1 & 2 & 3 & 4 & 5 & 6 & 7 & 8 & 9 & 10\\
\hline
$A_i$    & 660 & 670 & 700 & 680 & 450 & 970 & 480 & 665 & 1000 & 370\\
$B_i$    & 25.92 & 27.79 & 16.6 & 16.5 & 19.7 & 17.26 & 27.74 & 27.27 & 16.19 & 22.26\\
$C_i$    & 0.00413 & 0.00173 & 0.002 & 0.00211 & 0.00398 & 0.00031 & 0.0079 & 0.00222 & 0.00048 & 0.00712\\
$P_i^{min}$ & 10 & 10 & 20 & 20 & 25 & 150 & 25 & 10 & 150 & 20\\
$P_i^{max}$ & 55 & 55 & 130 & 130 & 162 & 455 & 85 & 55 & 455 & 80\\
\hline
\end{tabular}
\end{table*}
%%%%%%%%%%%%%%%%%%%%%%%%%%%%%%%%%%%%%%%%%%%%%%%%%%%%%%%%%%
%%%%%%%%%%%%%%%%%%%%%%%%%%%%%%%%%%%%%%%%%%%%%%%%%%%%%%%%%%%
\begin{table*}[ht]
\caption{Unit Commitment Results}
\label{table:2}
\centering
\begin{tabular}{p{0.1\linewidth}p{0.05\linewidth}p{0.05\linewidth}p{0.06\linewidth}p{0.06\linewidth}p{0.06\linewidth}p{0.05\linewidth}p{0.06\linewidth}p{0.06\linewidth}p{0.04\linewidth}p{0.06\linewidth}}
\hline
Load/Units & 1 & 2 & 3 & 4 & 5 & 6 & 7 & 8 & 9 & 10\\
\hline
%70    & On & On & On &  &  &  &  & On &  & On\\
100   & on & on &  & & on &  & on & on &  & on\\
200   & on & on & on & on & on &  & on & on &  & on\\
400   & on & on & on & on & on &  & on & on &  & on\\
800   & on & on & on & on & on & on & on & on &  & on\\
1000  & on & on & on & on & on & on & on & on & on & on\\
\hline
\end{tabular}
\end{table*}
%%%%%%%%%%%%%%%%%%%%%%%%%%%%%%%%%%%%%%%%%%%%%%%%%%%%%%%%%%

The number of binary variables in the QUBO problem is as many as units. The initialization step of Algorithm 1 plays a significant role in its convergence to the global optimum. Different load levels need different ranges of initialization. For load levels less than 100 MW, we set the range of $\rho$ and $\beta$ to around $ 10^6$. For load levels between 100 and 200 MW, we set $\rho=1001$ and $\beta=1000$, and for load levels greater than 200 MW, we set $\rho=4000$ and $\beta=1000$. The ADMM convergence tolerance $\epsilon$ is set to $10^{-6}$. Using Algorithm 1, with initialization of the same parameters, $S1$ and $S2$  obtain the same results as the base case.

To analyze the impact of updating variational parameters in the QAOA algorithm, we implemented $S2$ with no updating variational parameters. Fig. 2 depicts the ADMM residual, defined as $R^{\nu}=\sum_i ||y_i^{\nu}-z_i^{\nu}+r_i^{\nu}||$, for both $S1$ and $S2$ when load level is 800 MW. The residual's magnitude for both $S1$ and $S2$ is drawn for 50 iterations. In this instance, the optimal bitstring result has to be $|1011111111\rangle$, like what presented in Table \ref{table:2}, however, QAOA provides $|1111111111\rangle$. Fig. 3 illustrates the ADMM residual, once the variational parameters are updated in the QAOA algorithm. In this instance, the QAOA algorithm produces the same results as the classical solver.

%%%%%%%%%%%%%%%%%%%%%%%%%%%%%%%%%%%%%%%%%%%%%%%%%%%%%%%%%%%%%%%%%%%%%%%%%%%%
\begin{figure}
\centering
    \includegraphics[width=.48\textwidth]{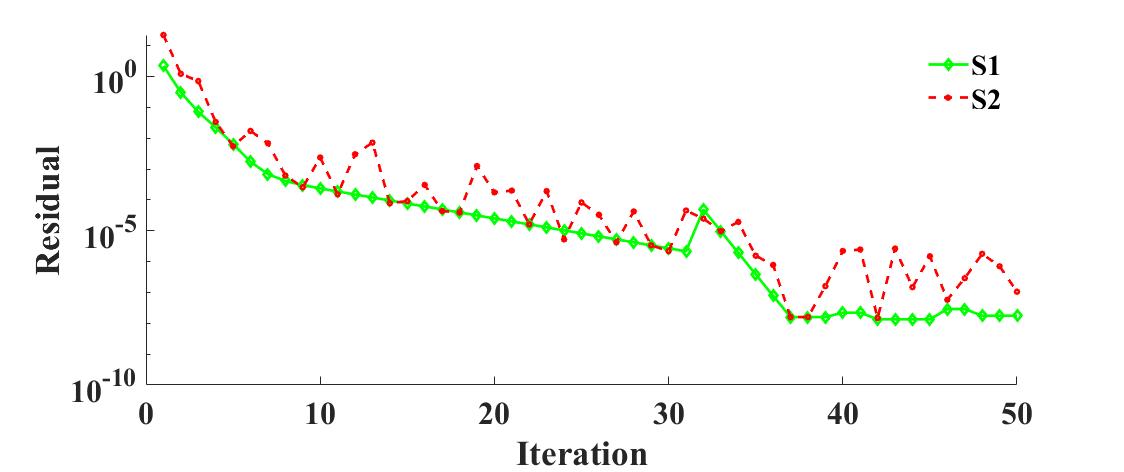}
\caption{ADMM Residual, not updating variational parameters in $S2$.}
\label{fig2}
\end{figure}
%%%%%%%%%%%%%%%%%%%%%%%%%%%%%%%%%%%%%%%%%%%%%%%%%%%%%%%%%%%%%%%%%%%%%%%%%%%%
%%%%%%%%%%%%%%%%%%%%%%%%%%%%%%%%%%%%%%%%%%%%%%%%%%%%%%%%%%%%%%%%%%%%%%%%%%%%
\begin{figure}
\centering
    \includegraphics[width=.48\textwidth]{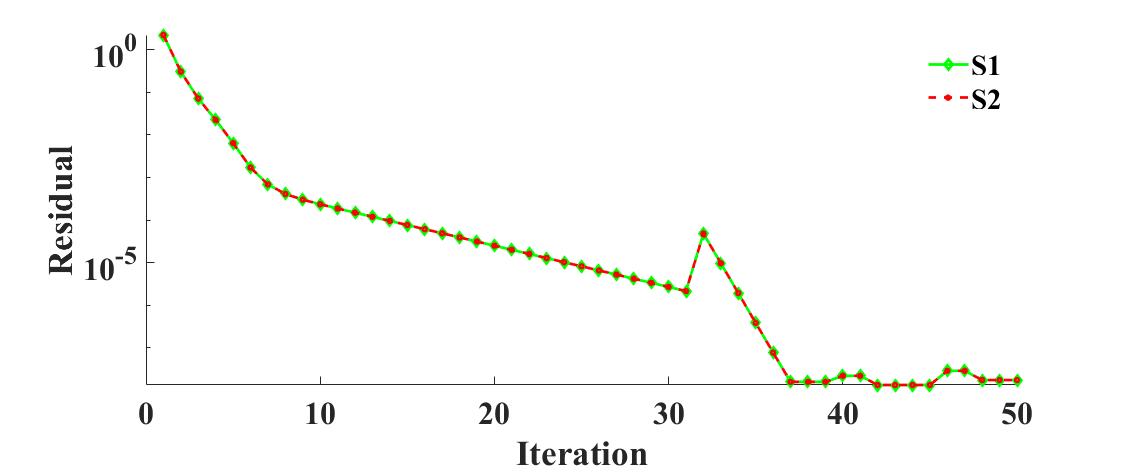}
\caption{ADMM Residual, with updating variational parameters in $S2$.}
\label{fig3}
\end{figure}
%%%%%%%%%%%%%%%%%%%%%%%%%%%%%%%%%%%%%%%%%%%%%%%%%%%%%%%%%%%%%%%%%%%%%%%%%%%%
To scrutinize the performance of QAOA, and the effect of the updating variational parameters on its performance, we have run $S2$ considering the first four units and 50 MW of load. The optimal solution of this case is $|11\rangle = |1011\rangle$, which means that units 1, 2, and 4 are on, and unit 3 is off. Fig. 4 illustrates the probability of the state of each bitstring in the last iteration of the ADMM. Two bitstrings $|3\rangle = |0011\rangle$ and $|11\rangle = |1011\rangle$ have the same probability before doing the measurement while we know only the bitstring $|11\rangle$ is the global optimum. In this case, we have run the QAOA algorithm with the best possible initialization guess for variational parameters. Once we update the variational parameters using previous $|\gamma,\beta \rangle$, the probability of achieving optimal bitstring increase as demonstrated in Fig. 5.

%%%%%%%%%%%%%%%%%%%%%%%%%%%%%%%%%%%%%%%%%%%%%%%%%%%%%%%%%%%%%%%%%%%%%%%%%%%%
%%%%%%%%%%%%%%%%%%%%%%%%%%%%%%%%%%%%%%%%%%%%%%%%%%%%%%%%%%%%%%%%%%%%%%%%%%%%
\begin{figure}
\centering
    \includegraphics[width=.48\textwidth]{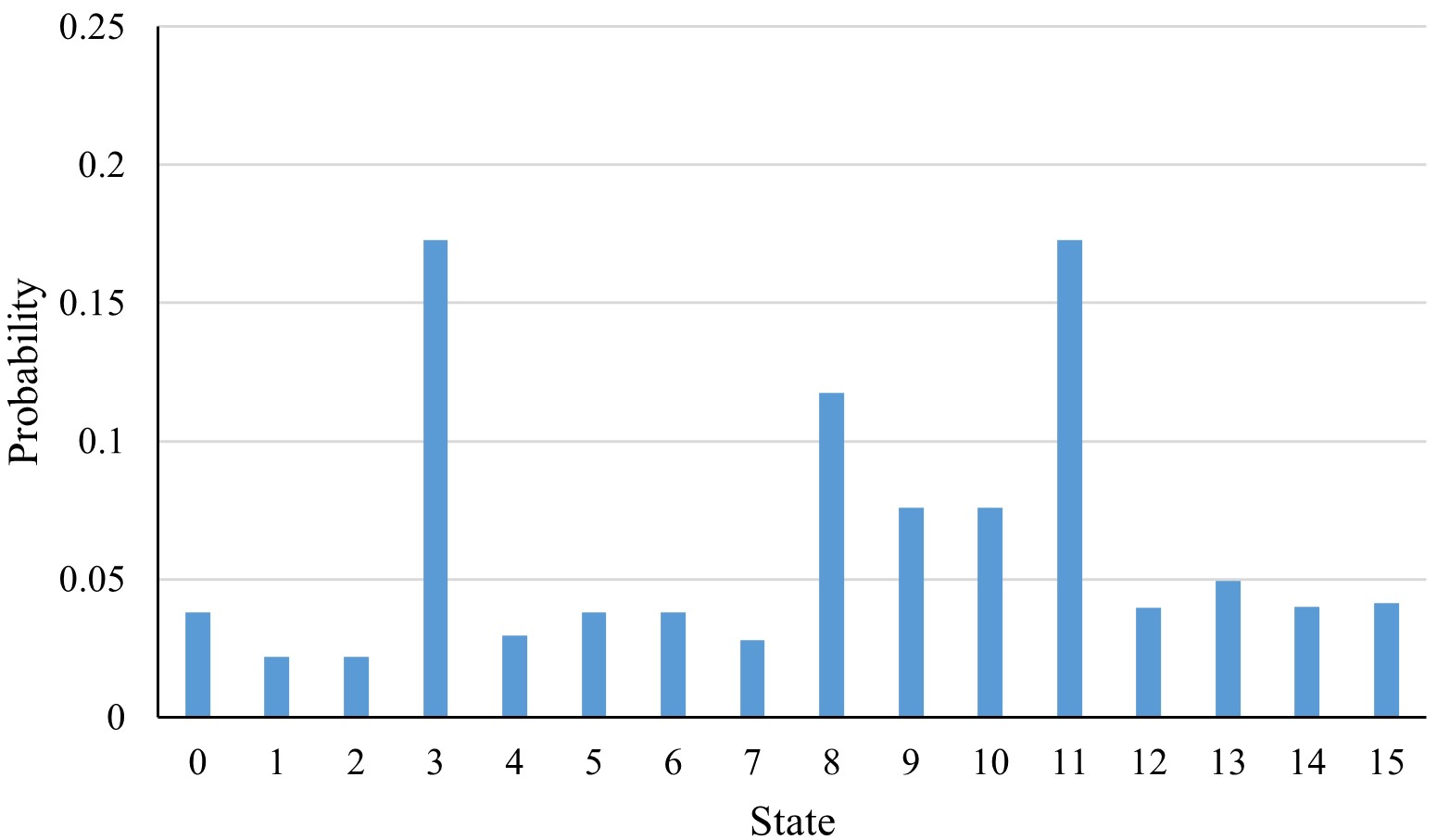}
\caption{The probability of each bitstring without updating the variational parameters.}
\label{fig4}
\end{figure}
%%%%%%%%%%%%%%%%%%%%%%%%%%%%%%%%%%%%%%%%%%%%%%%%%%%%%%%%%%%%%%%%%%%%%%%%%%%%
%%%%%%%%%%%%%%%%%%%%%%%%%%%%%%%%%%%%%%%%%%%%%%%%%%%%%%%%%%%%%%%%%%%%%%%%%%%%
\begin{figure}
\centering
    \includegraphics[width=.48\textwidth]{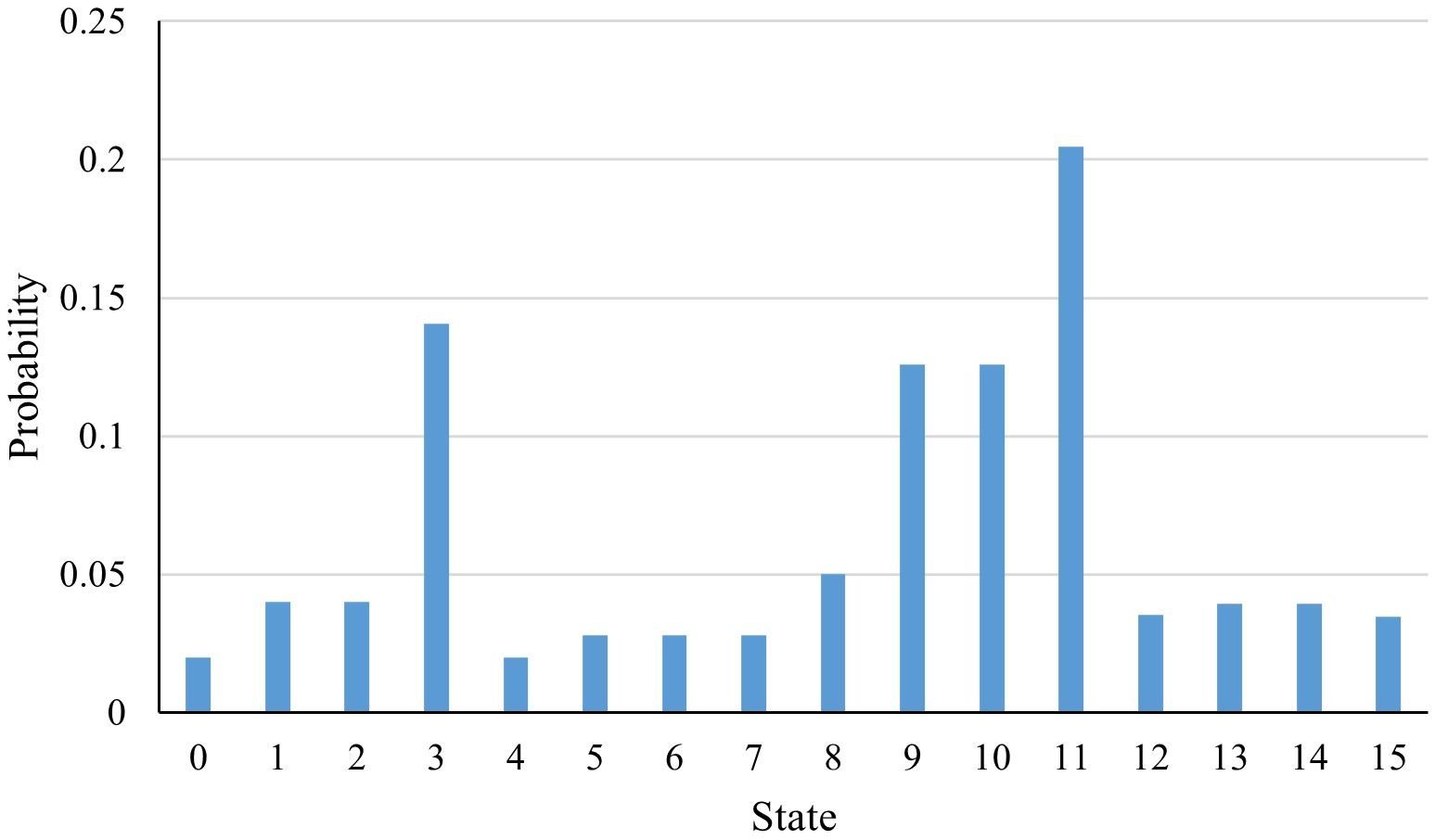}
\caption{The probability of each bitstring with updating the variational parameters.}
\label{fig5}
\end{figure}
%%%%%%%%%%%%%%%%%%%%%%%%%%%%%%%%%%%%%%%%%%%%%%%%%%%%%%%%%%%%%%%%%%%%%%%%%%%%
\section{Conclusion}
An algorithm is proposed to solve the UC problem using a combination of quantum and classical computers. A reformulation strategy is presented to decompose the problem into a QUBO and two non-QUBO subproblems. The non-QUBO subproblems are solved on classical computers, while the QUBO subproblem is solved using QAOA, which is a quantum computing algorithm. To obtain the optimal solution for the whole UC problem, the solution of subproblems are coordinated iteratively using a three-block ADMM. An updating strategy for QAOA variational parameters is applied to make this algorithm achieve the best possible results.

Simulation studies on a classical computer show the performance of the decomposition algorithm such that the results of the problem for different load levels match the results of a centralized classical solver. Moreover, when we solve the problem using the hybrid quantum-classical algorithm, we obtain the same results as those found by solving all subproblems in a classical computer.
%%%%%%%%%%%%%%%%%%%%%%%%%%%%%%%%%%%%%%%%%%%%%%%%%%%%%%%%%%%%%
\vspace{12pt}
\bibliographystyle{ieeetr}
\bibliography{Q-Classical-UC.bib}
%%%%%%%%%%%%%%%%%%%%%%%%%%%%%%%%%%%%%%%%%%%%%%%%%%%%%%%%%%%%%
\end{document}